\documentclass{iopart}
\usepackage{graphicx,amssymb} 

\begin{document}

\title{An edge index for the Quantum Spin-Hall effect}

\author{Emil Prodan}

\address{Department of Physics, Yeshiva University, New York, NY 10016}

\begin{abstract}
Quantum Spin-Hall systems are topological insulators displaying dissipationless spin currents flowing at the edges of the samples. In contradistinction to the Quantum Hall systems where the charge conductance of the edge modes is quantized, the spin conductance is not and it remained an open problem to find the observable whose edge current is quantized.  In this paper, we define a particular observable and the edge current corresponding to this observable. We show that this current is quantized and that the quantization is given by the index of a certain Fredholm operator. This provides a new topological invariant that is shown to take the generic values 0 and 2, in line with the Z$_2$ topological classification of time-reversal invariant systems. The result gives an effective tool for the investigation of the edge structure in Quantum Spin-Hall systems. Based on a reasonable assumption, we also show  that the edge conducting channels are not destroyed by a random edge. 
\end{abstract}


\maketitle

\section{Introduction} 

A new class of insulators has been recently found \cite{Kane:2005vn,Kane:2005ys, Sheng:2005dz,B.A.-Bernevig:2006zr} to
possess a dissipationless Quantum Spin-Hall effect. Describing the structure of the edge modes in these systems remains an interesting 
issue for both fundamental understanding and
potential applications of the Quantum Spin-Hall effect \cite{Fu:2006p190,Sheng:2006vn,Fukui:2007sf,T:2007yg,Moore:2007rr,Essin:wd,Qi:2008eu,Qi:2008tk,Fukui:2008fq}. It was argued in the literature that the initial Z$_2$ topological classification proposed in Ref.~\cite{Kane:2005vn} can be further refined to meet this purpuse. Not long ago,  Ref.~\cite{Sheng:2006vn} introduced a new bulk topological invariant, called spin-Chern number, which seemed to the contain more information about the edge structure. Later, however, it became clear that the spin-Chern number is unstable to deformations of the Hamiltonian system and that the only invariants for the spin-Hall effect are of Z$_2$ type, rather than integer type \cite{Fukui:2007sf,Moore:2007rr,Essin:wd,Qi:2008tk}. 

Despite of the above concentrated efforts, that actually lead to a deep understanding of the topological insulators, there are still a few open problems. First, it is not completely clear how to describe the topological phases for non-crystaline systems. Secondly, it is well known \cite{Kane:2005vn,Kane:2005ys, Sheng:2005dz,Sheng:2006vn} that the spin edge current is not quantized in the Quantum Spin-Hall systems and it remained an open problem to find an observable that has such a quantized edge current. It is worth mentioning that for the Integer Hall Effect, a complete, rigorous description of the edge states that goes beyond crystaline systems  was achieved only in 2001 by Kellendonk, Richter and Schulz-Baldes \cite{SchulzBaldes:2000p599}. The present paper was inspired by a later work of these authors \cite{Kellendonk:2004p597} which deals with the quantization of edge currents for half-plane continuous magnetic operators in the presence of weak random potential. The formalism was put into an abstract setting in Ref.~\cite{Prodan:2008oq}, which actually provided the guiding lines for the present paper. This general formalism was applied in Ref.~\cite{Prodan:2008ai} to a simpler problem, namely the quantization of edge currents in Chern insulators with rough edges. The technical estimates derived in this paper are important for the present analysis.

The present paper addresses both open issues mentioned above. Using the time-reversal invariance property of Spin-Hall systems, we define an observable and its corresponding current and we show that the expectation value (taken only over the spectrum in the insulating gap) of this current is quantized and that the quantization is described by the index of a Fredholm operator. This is our new topological invariant, which we call the edge index. For the model considered in Ref.~\cite{Sheng:2006vn}, we show that this invariant takes the same value as the Spin-Chern number defined in same reference. For the general case, we show that the edge index takes the generic values 0 or 2, in line with the Z$_2$ classification of the time-reversal invariant insulators.

\begin{figure}
 \center
 \includegraphics[width=7cm]{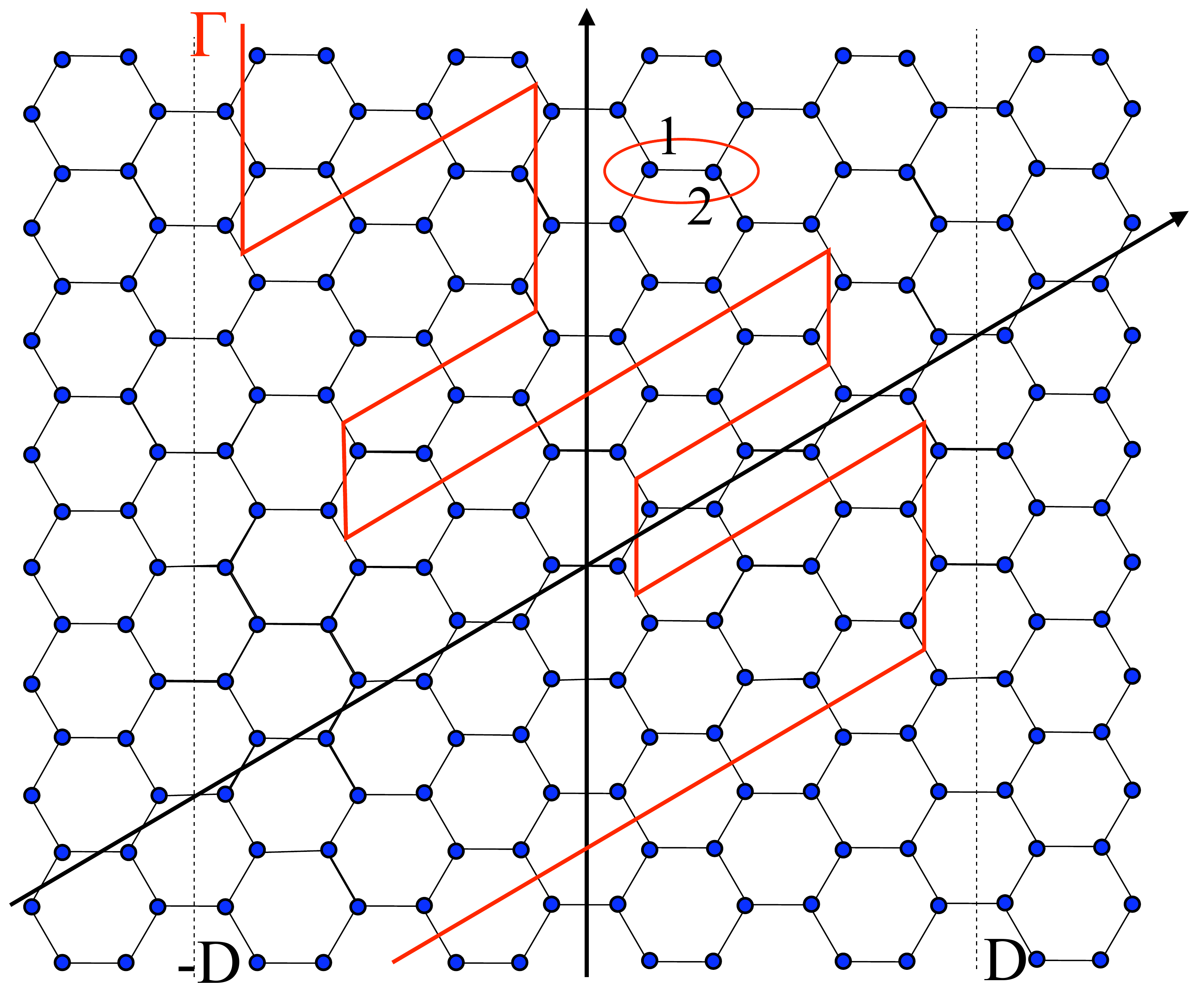}
 \caption{The figure illustrates the honeycomb lattice, an example of paired sites with the index $a$ attached to each site, and a random edge $\Gamma$. The contour $\Gamma$ never crosses the bonds between the pairs and is contained between the vertical lines at $-D$ and $D$.}
 \label{division}
\end{figure}

\section{The model}

\begin{figure}
 \center
 \includegraphics[width=6cm]{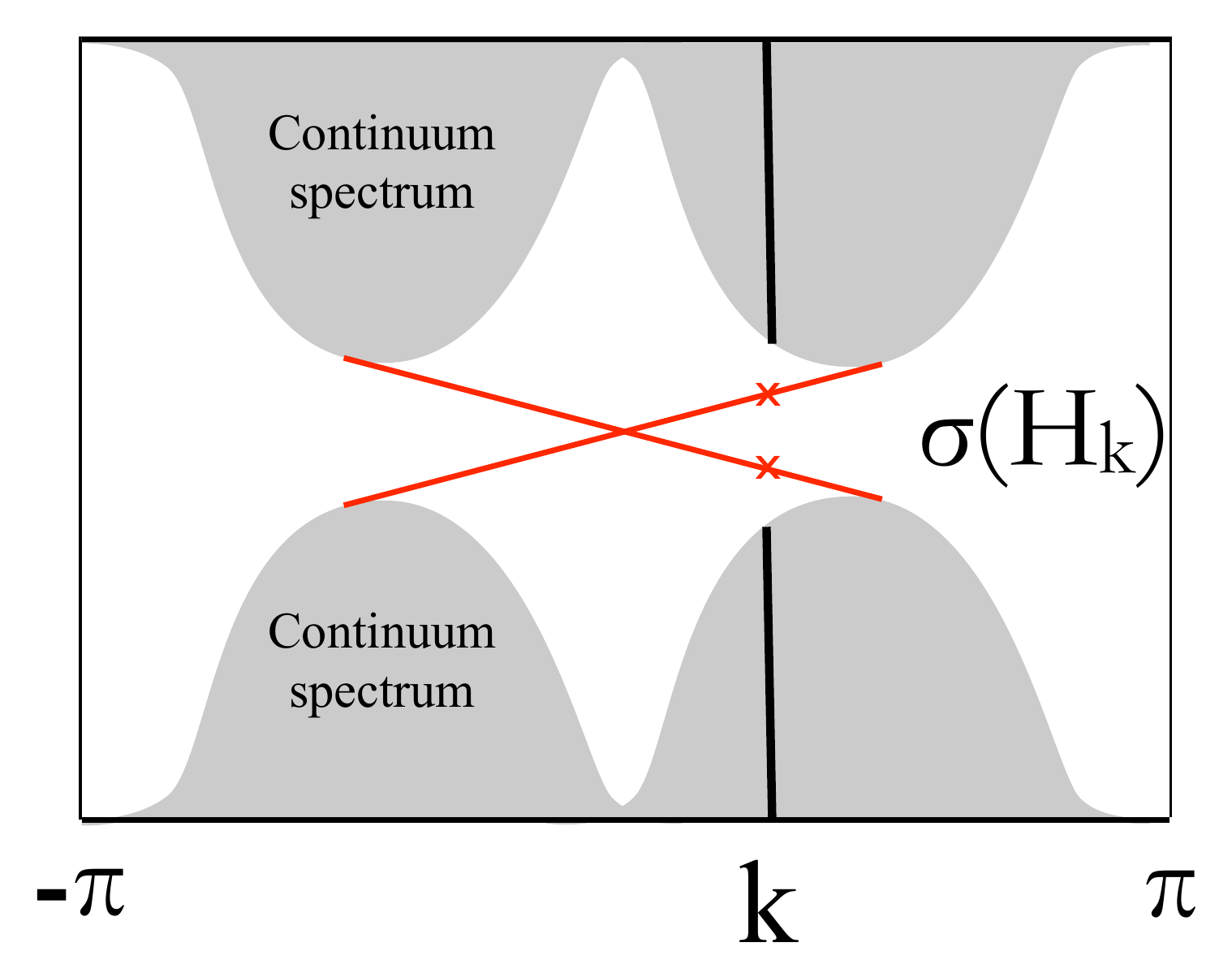}
 \caption{The figure illustrates the energy spectrum of the Bloch Hamiltonians $H_k$ corresponding to the Bloch decomposition of the edge Hamiltonian relative to the translational symmetry along the homogenous edge.}
\end{figure}

To be concrete, we consider non-interacting electrons on a honeycomb lattice (see Fig. 1) described by the bulk Hamiltonian of Ref.~\cite{Sheng:2006vn}:
\begin{equation}\label{h1}
\begin{array}{c}
H_0=-t\sum\limits_{\langle ij \rangle,\alpha} |i,\alpha\rangle \langle j,\alpha| \medskip \\
+iV_{SO}\sum\limits_{\langle \langle ij \rangle \rangle,\alpha\beta}  [ {\bf \sigma} \cdot ({\bf d}_{kj} \times {\bf d}_{ik} )]_{\alpha,\beta} | i,\alpha\rangle \langle  j,\beta| \medskip \\
+iV_R\sum\limits_{\langle \langle ij \rangle \rangle,\alpha\beta}  [ \hat{{\bf z}}\cdot ({\bf \sigma}\times {\bf d}_{ij})]_{\alpha,\beta} |i,\alpha\rangle \langle j,\beta|.
\end{array}
\end{equation}
This particular model does not play any critical role in our analysis, except that it displays all the general features that we mention in the following. The Hamiltonian of Eq.~\ref{h1} has time reversal symmetry and is a good model for electrons in graphene \cite{Kane:2005ys}. The first term is the usual nearest neighbor hopping term, the second term is an intrinsic SO coupling preserving the lattice symmetries and the third term stands for the Rashba SO coupling. For details about the notation please consult Ref.~\cite{Sheng:2006vn}. In the following, we consider that we are in the Spin-Hall part of the phase diagram of the model \cite{Kane:2005vn}.

The bulk model displays two  top bands and two bottom bands separated by a gap. The two bottom bands have opposite Chern numbers $c$=$\pm 1$, so their total Chern number is zero. When $V_R$=0, $S_z$ commutes with the Hamiltonian and the model Eq.~\ref{h1} reduces to a spin up and a spin down decoupled Haldane models \cite{Haldane:1988dz}. In contradistinction to the Chern numebr, the Spin-Chern number $c_s$ introduced in Ref.~\cite{Sheng:2006vn} is nontrivial:  if $V_R$=0, it reduces to $c_{s}= c_{\uparrow}-c_{\downarrow}$ (=$\pm$2 for the model Eq.~\ref{h1}, depending on the sign of $V_{SO}$). $c_s$ can be generalized to the case when $S_z$ is not conserved, like when the Rashba term is present. After extensive numerical analysis, Ref.~\cite{Sheng:2006vn} concluded that the Spin-Chern number remains quantized when $V_R$ and a weak disorder are turned on. 

The special topological properties of the bulk energy bands have non-trivial consequences for the surface states spectrum when an edge is cut on a bulk sample. Let us briefly discuss the edge spectrum for a homogeneous edge. In this case we can use the Bloch decomposition with respect to the periodicity along the edge and write the edge Hamiltonian as a continuous direct sum of Bloch Hamiltonians $H_k$. As illustrated in Fig.~2, the spectrum of each $H_k$ consists of upper and lower continuum parts plus two nondegenerate (excepting $k$=0), discrete eigenvalues. These discrete eigenvalues for different $k$'s assemble themselves in two bands, shown in red color in Fig.~2. If the Rashba term is zero, one band corresponds to the spin up and the other band to the spin down. Thus, while the charge moves in opposite directions for these two bands (leading to zero charge current), the spins move in the same direction and consequently the edge carries a dissipationless spin current. The edge modes are protected by the time reversal symmetry, which means no gap can open in the edge spectrum, even when the Rashba term is turned on. While $S_z$ is no longer conserved for this later case, the edge still carries a dissipationless spin current, thought no longer quantized. Because of the last fact, the theory of Quantum Spin-Hall is still missing a topological invariant that could tell how many edge bands one should expect in more complicated models. Finding such an invariant is the goal of the present paper. 

Our analysis will be done on an equivalent system, a triangular lattice with 4 quantum states per site. This system is obtained by considering the honeycomb lattice as composed of pairs of sites sitting on a triangular lattice. For example, the 4 quantum states residing on the pair of sites circled in Fig. 1 can be thought as 4 quantum states residing at a new lattice site positioned at the mid point between the pair. This way we obtain an equivalent triangular lattice model with 4 quantum states per site (see Fig. 2). The Hilbert space is now spanned by the states:
\begin{equation}
|{\bf n},{\bf a}\rangle, {\bf n}=(n_1,n_2) \in \mbox{triangular lattice}, \ {\bf a}=(a,\alpha),
\end{equation}
where $a$=1,2 is the index introduced in Fig.~1 and $\alpha$ is the spin index. The triangular lattice sites are described by $(n_1,n_2)$, where $n_1$ and $n_2$ represent the coordinates along the two directions shown in Fig.~2. The bulk Hamiltonian becomes:
\begin{equation}\label{h2}
H_0=\sum_{{\bf n},{\bf n}'} \sum_{{\bf a},{\bf b}}[\Gamma_{{\bf a} {\bf b}}^{{\bf n}{\bf n}'} |{\bf n},{\bf a}\rangle \langle  {\bf n}',{\bf b} | + \bar{\Gamma}_{{\bf a} {\bf b}}^{{\bf n}{\bf n}'} |{\bf n}',{\bf b}\rangle \langle  {\bf n},{\bf a} |].
\end{equation}
The coefficients $\Gamma_{{\bf a} {\bf b}}^{{\bf n}{\bf n}'}$ can be computed from Eq.~\ref{h1}, but their explicit expression is not needed here. The first sum is over the nearest neighbors.

\begin{figure}
 \center
 \includegraphics[width=7cm]{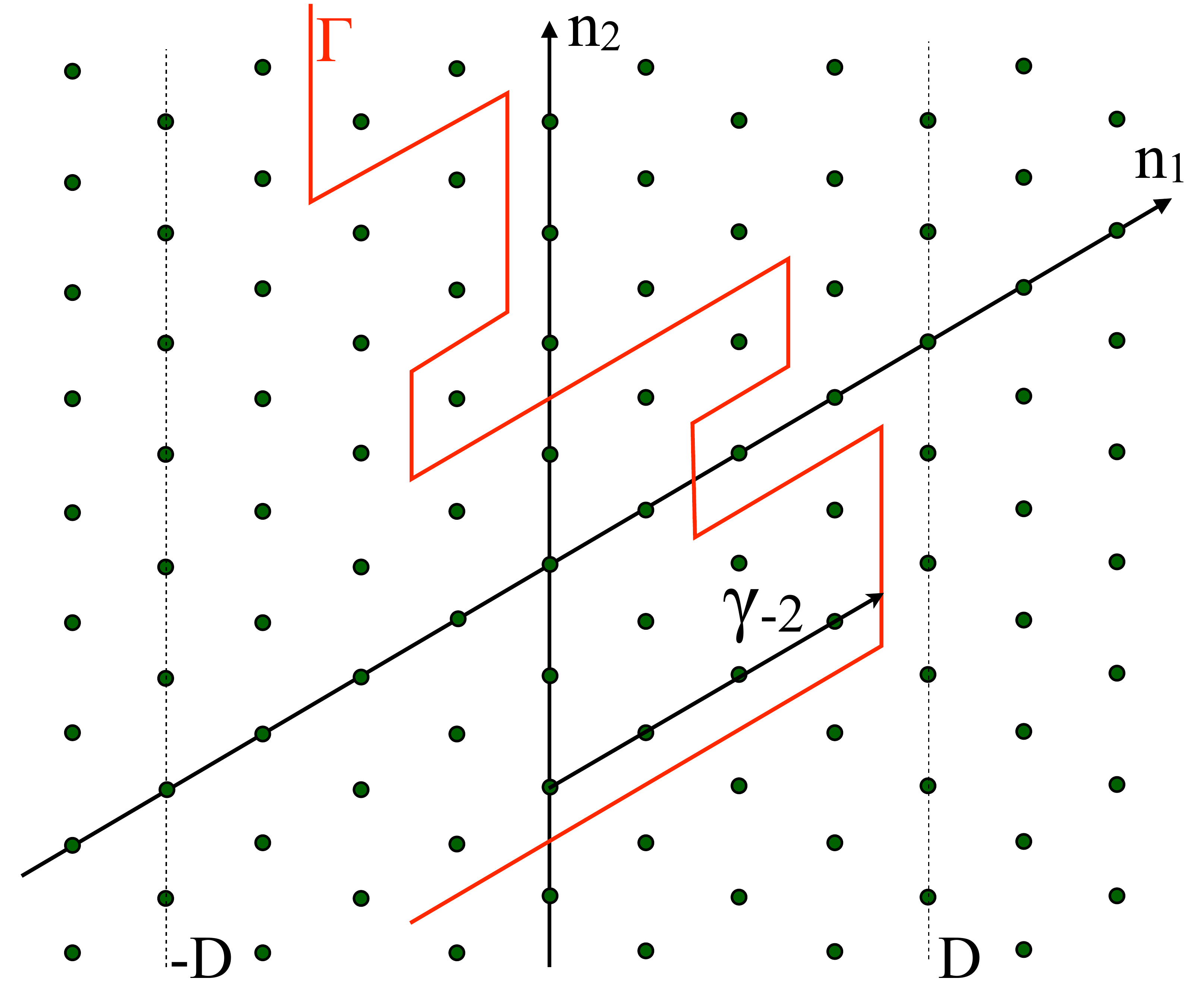}
 \caption{The figure illustrates the equivalent triangular lattice. It also shows the contour $\Gamma$ from Fig.~1. This $\Gamma$ can be described by a sequence $\{\gamma_n\}_n$ which gives the deviation of $\Gamma$ from the $n_1$=0 axis, at $n_2$=$n$. The figure illustrates how $\gamma_{-2}$ is defined.}.
\end{figure}

We now consider the system with the edge. In the lab samples, the strongest irregularities are probably seen at the edges of the samples, so here we will concentrate at this type of disorder and we will neglect the bulk disorder (the mathematics still work for weak bulk disorder). On the honeycomb lattice we consider random contours $\Gamma$, like the one shown in Fig.~1, their main features being that they never cross the bond between the pairs and that they are confined within $-D<n_1<D$, where $D$ will be fixed from now on. On the triangular lattice, $\Gamma$ can be described by a sequence $\{\gamma_n\}_n$, where $\gamma_n$ gives the deviation of $\Gamma$ from the axis $n_1$=0 at the row $n_2$=$n$ of the lattice, as illustrated in Fig.~2. We have $\gamma_n \in {\cal I}$, with ${\cal I}=\{-D+1/2,-D+3/2, \ldots,D-1/2\}$.  Thus, $\Gamma$ can be viewed as a point of the set $\Omega={\cal I}^{\times \infty}$: $\Gamma$=$\{\ldots,\gamma_{-1},\gamma_0,\gamma_1,\ldots\}$. On the set $\Omega$, we introduce the product probability measure, denoted by $d\Gamma$, which is the infinite product of the simplest probability measure $\nu$ on ${\cal I}$: $\int f(n) d\nu(n)=\frac{1}{2D}\sum_{n\in {\cal I}} f(n)$, $f(n)$ being any function defined on ${\cal I}$. We remark that $d\Gamma$ obtained in this way is ergodic relative to the discrete translations along the vertical direction of our lattice. We will use $d\Gamma$ to average over all possible contours $\Gamma$.

The system with the edge is defined on the Hilbert space
${\cal H}_\Gamma =span\{ |{\bf n},{\bf a}\rangle, \ {\bf n} \ \mbox{to the right of} \ \Gamma\}$ and its Hamiltonian is given by $H_\Gamma : {\cal H}_\Gamma \rightarrow {\cal H}_\Gamma$,
\begin{equation}\label{h2}
H_\Gamma=\sum_{{\bf n},{\bf n}'} \sum_{{\bf a},{\bf b}}[\Gamma_{{\bf a} {\bf b}}^{{\bf n}{\bf n}'} |{\bf n},{\bf a}\rangle \langle  {\bf n}',{\bf b} | + \bar{\Gamma}_{{\bf a} {\bf b}}^{{\bf n}{\bf n}'} |{\bf n}',{\bf b}\rangle \langle  {\bf n},{\bf a} |],
\end{equation}
where the first sum is restricted to the sites located to the right of $\Gamma$. $H_\Gamma$ remains time reversal invariant.

\section{The main result}

We define now the central observable.  As it was pointed out for the case of translational, time reversal invariant, half-integer spin Hamiltonians \cite{Kane:2005vn,Kane:2005ys,Fu:2006p190}, the Hilbert space can be divided in two invariant subspaces. This remains true when the translational symmetry is broken. More precisely, the Hilbert spaces ${\cal H}_\Gamma$ can be decomposed as ${\cal H}_\Gamma={\cal H}_-(\Gamma) \oplus {\cal H}_+(\Gamma)$, where the orthogonal subspaces ${\cal H}_\pm(\Gamma)$ have the following special properties:
\begin{equation}\label{property}
\theta {\cal H}_\pm(\Gamma) = {\cal H}_\mp(\Gamma) \ \mbox {and} \ H_\Gamma {\cal H}_\pm(\Gamma) ={\cal H}_\pm(\Gamma), 
\end{equation}
where $\theta$ denotes the time reversal operation, $\theta = e^{i\pi S_y/\hbar}K$ ($K$= complex conjugation). An important observation here is that the construction is not unique. Let us denote by $\Pi_\pm^i$ the orthogonal projectors onto ${\cal H}_\pm(\Gamma)$ and define $\Sigma^i_\Gamma \equiv \Pi_+^i - \Pi_-^i$, where we reintroduced the index $\Gamma$ to remind that the operator is defined on ${\cal H}_\Gamma$. Our central observable is defined by the self-adjoint operator:
\begin{equation}
X_\Gamma =\frac{1}{2} (y_\Gamma \Sigma_\Gamma^i+\Sigma_\Gamma^i y_\Gamma),
\end{equation}
where $y_\Gamma|{\bf n},\alpha \rangle = n_2 |{\bf n},\alpha \rangle$, defined on ${\cal H}_\Gamma$, is the observable giving the vertical coordinate. The self-adjoint property of the central observable can be demonstrated by following a technique developed in Ref.~\cite{Nenciu:1998cr}.

Our systems with edge and the observables $X_\Gamma$ have very special properties under vertical translations of the lattice. Let
\begin{equation}
u_n | (n_1,n_2),{\bf a} \rangle = | (n_1,n_2-n),{\bf a} \rangle.
\end{equation}
be the implementation of the lattice translations along the $n_2$ direction. These translations can also be extended to a map $t_n$ acting on the space $\Omega$ of all possible contours $\Gamma$. The map $t_n$ simply shifts a contour downwards by $n$ sites. We now can list those special properties:
\begin{enumerate}
\item The family $\{H_\Gamma\}_{\Gamma \in \Omega}$ is covariant: $u_n H_\Gamma u_n^* = H_{t_n \Gamma}$.
\item Based on 1, we can choose $\Sigma_\Gamma^i$ such that $u_n \Sigma_\Gamma^i u_n^* = \Sigma_{t_n \Gamma}^i$. Moreover, $[\Sigma_\Gamma^i,H_\Gamma]=0$.
\item The central observable obeys: 
\begin{equation}
u_n X_\Gamma u_n^* =X_{t_n \Gamma} + n \Sigma_{t_n \Gamma}^i, \ [X_\Gamma,\Sigma_\Gamma^i]=0.
\end{equation} 
\item For any function $f(\epsilon)$, commutators of the form $[X_\Gamma,f(H_\Gamma)]$ form covariant families :
\begin{equation}
u_n[X_\Gamma,f(H_\Gamma)]u_n^* =   [X_{t_n \Gamma},f(H_{t_n \Gamma})].
\end{equation}
\end{enumerate}

 We are now gearing towards the main result. We denote the spectral projector of $X_\Gamma$ onto the spectrum inside interval $[n$$-$$1/2,n$+$1/2)$ by $\pi_\Gamma(n)$. Note that, at least for a small Rashba term, the half-integer numbers are outside the eigenvalue spectrum of $X_\Gamma$. This can be shown via estimates on the resolvent of $X_\Gamma$ using techniques developed in Ref.~\cite{Nenciu:1998cr}. If $\mbox{tr}_0 A\equiv \mbox{Tr} \{\pi_\Gamma(0) A \pi_\Gamma(0)\}$, we define the current of $X_\Gamma$ as:\cite{Prodan:2008oq}
\begin{equation}
J_\Gamma =  \mbox{tr}_0 \left\{\rho(H_\Gamma) \frac{\mbox{d}X_\Gamma(t)}{\mbox{d}t}\right\} =i \mbox{tr}_0 \left\{\rho(H_\Gamma) [H_\Gamma,X_\Gamma ]\right\}.
\end{equation}
Here $\rho(\epsilon)$ is the statistical distribution of the quantum states. Since we are interested in the contributions from the edge states, we assume that $\rho(\epsilon)$ is a smooth function with support in the bulk insulating gap. 

Tight-binding Hamiltonians like $H_\Gamma$ were analyzed in Ref.~\cite{Prodan:2008ai}. With the {\it assumption} that the amplitude of $\pi_\Gamma(0)| {\bf n},{\bf a} \rangle$ decays sufficiently fast for large $|n_2|$, the technical estimates given in Ref.~\cite{Prodan:2008ai} assure that, in the present article, all the operators appearing inside the traces are trace class (so the trace is finite and independent of the basis set used to compute it) and all the sums are absolutely convergent.

 {\bf Main Statement.}  Let $F(\epsilon)\equiv \int_{\epsilon}^{\infty} \rho(\epsilon)$. Note that $F(\epsilon)$ is smooth and equal to 1/0 below/above the bulk insulating gap; also $F'(\epsilon)$=$-\rho(\epsilon)$. We define the following unitary operators: $U_\Gamma = e^{-2 \pi i F(H_\Gamma)}$. If $\pi_\Gamma^>$ is the projector onto the non-negative spectrum of $X_\Gamma$, then:
\begin{equation}\label{main}
\int_\Omega d\Gamma \ J_\Gamma  = \frac{1}{2\pi}   \ \mbox{Ind} \left \{\pi_\Gamma^> U_\Gamma \pi_\Gamma^> \right \}.
\end{equation} 

This is our main statement. Let us comment on it first. The index is an integer number, defined on the class of Fredholm operators as:
\begin{equation}
\mbox{Ind} A=\dim  Ker[A] -\dim  Ker[A^*].
\end{equation}
It has very special properties, the most important being the invariance to norm-continuous deformations of the operator that keep the operator inside the Fredholm class. In our case, it follows from the estimates of Ref.~\cite{Prodan:2008ai} that, as long as the the gap remains opened and the support of $\rho(\epsilon)$ remains inside the gap, we can deform $\rho(\epsilon)$ or $H_\Gamma$ without changing the index. Moreover, the index is  independent of the contour $\Gamma$. To see this, we turn off the Rashba term (without changing the index) and reduce the system to two decoupled Chern insulators. But for Chern insulators, it was already shown in Ref.~\cite{Prodan:2008ai} that the index is independent of contour $\Gamma$.

We now show that the index is equal to the Spin-Chern number introduced in Ref.~\cite{Sheng:2006vn}. We take $\Gamma$ as a straight vertical line. Without changing the index, we can turn the Rashba term to zero. In this case the up and down spins decouple and we can take ${\cal H}_\pm$ as the spin up and spin down invariant subspaces, respectively. Definitely Eq.~\ref{main} applies equally well to the case when the set $\Omega$ reduces to one point, the straight contour $\Gamma_0$ (all we have to do is to take $D$=0). Then we have the following practical way of computing the index:
\begin{equation}\label{index}
\begin{array}{c}
\mbox{Ind} \{\pi_{\Gamma_0}^> U_{\Gamma_0} \pi_{\Gamma_0}^> \}= i\mbox{tr}_0 \{  \rho(H_{\Gamma_0}) [H_{\Gamma_0},X_{\Gamma_0}]  \} \medskip \\
=i \mbox{Tr}_\uparrow \{ \pi_{\Gamma_0}(0) \rho(H_{\Gamma_0}) [H_{\Gamma_0},y_{\Gamma_0}] \pi_{\Gamma_0}(0) \} \medskip \\
-i\mbox{Tr}_\downarrow \{ \pi_{\Gamma_0}(0) \rho(H_{\Gamma_0}) [H_{\Gamma_0},y_{\Gamma_0}] \pi_{\Gamma_0}(0) \} 
\end{array}
\end{equation}
Using the Bloch decomposition, this becomes
\begin{equation}
\sum\limits_n  \int\limits_{k=-\pi}^\pi [ \rho(\epsilon^\uparrow_{n k}) \partial_k \epsilon^\uparrow_{n k}  -  \rho(\epsilon^\downarrow_{n k}) \partial_k \epsilon^\downarrow_{n k} ] \ dk ,
\end{equation}
where $\epsilon^{\uparrow,\downarrow}_{n,k}$ are the edge energy bands. Since $\int \rho(\epsilon)=1$, each integral gives the difference between the number of forward and backward moving bands for the corresponding spin, known to equal the Chern number for the corresponding spin. Thus, the index is equal to the difference between the Chern numbers for spin up and spin down, i.e. it takes the same value as the Spin-Chern number introduced in Ref.~\cite{Sheng:2006vn}.

Note that our main statement is about the average of the edge current and not the current itself. However, since the family $\{H_\Gamma\}_{\Gamma \in \Omega}$ is covariant relative to translations, which act ergodically on $\Omega$,  the spectrum of $H_\Gamma$ is non-random. This implies that, if the edge spectrum becomes localized for a non-zero measure subset of $\Omega$, it will be localized for all contours, except a possible zero measure subset of $\Omega$. But this cannot happen, exactly because the average of the edge current is non-zero for Spin-Hall insulator. This allows us to conclude that the rough edge cannot destroy the edge conducting channels.

\begin{figure}
\center
 \includegraphics[width=7cm]{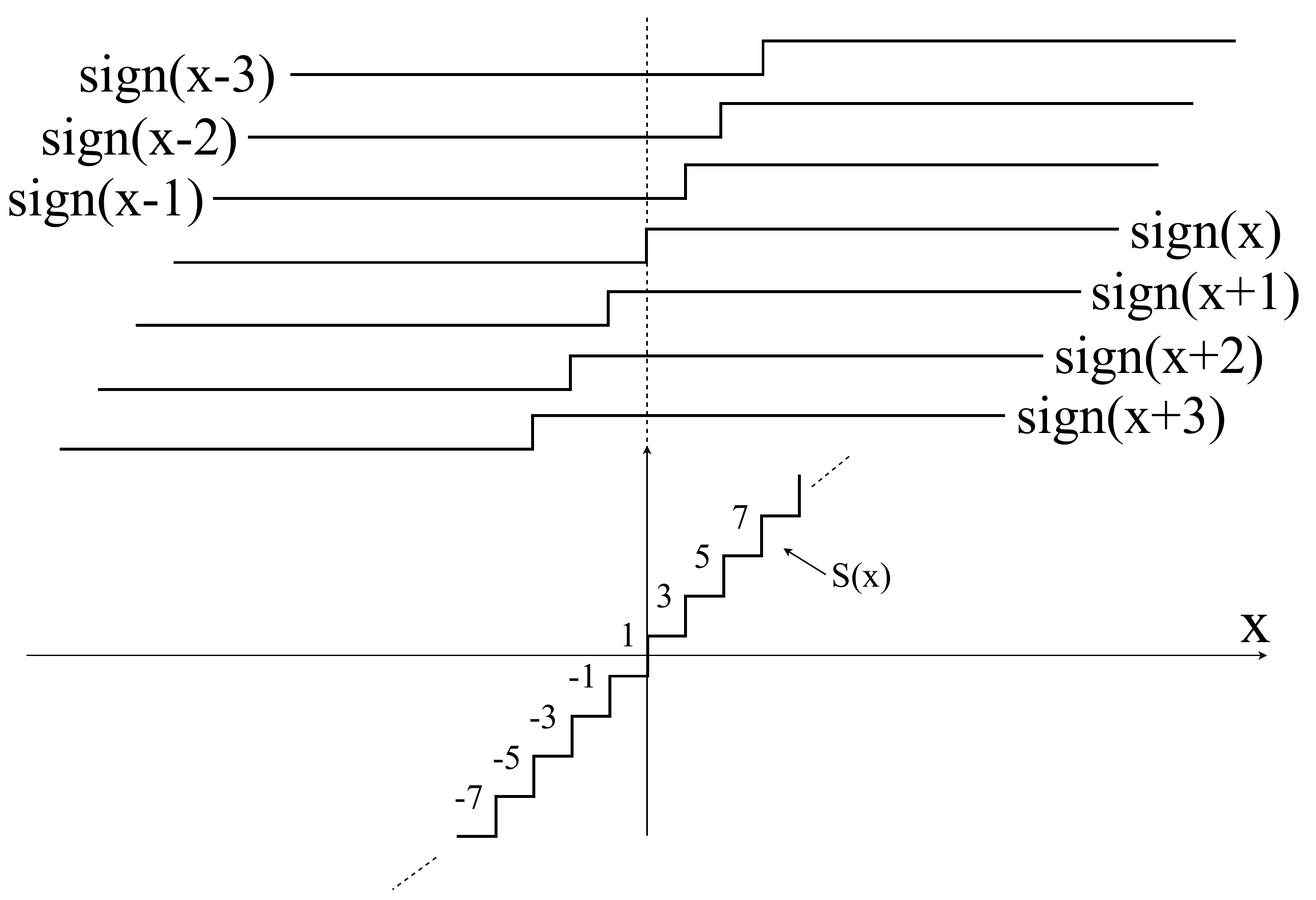}\\
 \caption{A graphical representation of $\sum_n \mbox{sign}(x+n)=S(x)$. The top lines represent the shifted sign functions $\mbox{sign}(x+n)$. The sum of the top lines results in the stair like function $S(x)$ represented by the bottom line.}  
 \label{fig4}
 \end{figure}

\section{Sketch of Proof} 

With our assumption that the amplitude of $\pi_\Gamma(0)| {\bf n},{\bf a} \rangle$ decays sufficiently fast for large $|n_2|$, it follows from the technical estimates of Ref.~\cite{Prodan:2008ai} that $\pi_\Gamma^> U_\Gamma \pi_\Gamma^>$ is in the Fredholm class. Let $\pi_\Gamma^<$ be the projector onto the negative spectrum of $X_\Gamma$ and $\Sigma_\Gamma\equiv \pi_\Gamma^> - \pi_\Gamma^<$. We compute the index using the formula:\cite{Prodan:2008oq,Prodan:2008ai} 
\begin{equation}\label{expand}
\begin{array}{c}
\mbox{Ind}\{\pi_\Gamma^> U_\Gamma \pi_\Gamma^>\} \medskip \\
=-\frac{1}{2}\sum\limits_{n} \mbox{Tr}\{\pi_\Gamma(n) (U_\Gamma^*-I)[\Sigma_\Gamma,U_\Gamma]\pi_\Gamma(n)\} ,
\end{array}
\end{equation}
where the sum is absolutely convergent. The projectors $\pi_\Gamma(n)$ leave the subspaces ${\cal H}_\pm (\Gamma)$ invariant, so they decompose in a direct sum: $\pi_\Gamma(n) = \pi_\Gamma^-(n)\oplus \pi_\Gamma^+(n)$. Similarly for $\Sigma_\Gamma$: $\Sigma_\Gamma = \Sigma_\Gamma^- \oplus \Sigma_\Gamma^+$. Due to property (3) listed above we have the following fact:
\begin{equation}
u_n \pi_\Gamma^\pm(m) u_n^* = \pi_{t_n \Gamma}^\pm (m\mp n).
\end{equation} 
We consider now the average over $\Gamma$. Since the index is  independent of $\Gamma$, the operation can be omitted for the left hand side. On the right hand side, we use the fact that the trace of trace-class operators is invariant to unitary transformations and that the measure $d\Gamma$ is invariant to the mappings $t_n$, to write:
\begin{equation}\label{step1}
\begin{array}{c}
-2 \mbox{Ind}\{\pi_\Gamma^> U_\Gamma \pi_\Gamma^>\}  \medskip \\
 = \sum\limits_{n} \int d\Gamma \ \mbox{Tr}\{u_{-n}\pi_\Gamma^-(n) (U_\Gamma^*-I)[\Sigma_\Gamma,U_\Gamma]\pi_\Gamma^-(n) u_{-n}^* \} \medskip \\
 +\sum\limits_{n}   \int d\Gamma \ \mbox{Tr}\{u_{n}\pi_\Gamma^+(n) (U_\Gamma^*-I)[\Sigma_\Gamma,U_\Gamma]\pi_\Gamma^+(n) u_{n}^* \} \medskip \\
=\sum\limits_{n} \int d\Gamma \times \medskip \\
( \mbox{Tr}\{\pi_{t_{-n}\Gamma}^-(0) (U_{t_{-n}\Gamma}^*-I)[u_{-n}\Sigma_\Gamma^- u_{-n}^*,U_{t_{-n}\Gamma}]\pi_{t_{-n}\Gamma}^-(0) \} \medskip \\
+\mbox{Tr}\{\pi_{t_{n}\Gamma}^+(0) (U_{t_n\Gamma}^*-I)[u_{n}\Sigma_\Gamma^+ u_{n}^*,U_{t_n\Gamma}]\pi_{t_{n}\Gamma}^+(0) \}  ) \medskip \\
=\sum\limits_{n} \int d\Gamma 
( \mbox{Tr}\{\pi_{\Gamma}^-(0) (U_{\Gamma}^*-I)[u_{-n}\Sigma_{t_n\Gamma}^- u_{-n}^*,U_{\Gamma}]\pi_{\Gamma}^-(0) \} \medskip \\
+\mbox{Tr}\{\pi_{\Gamma}^+(0) (U_{\Gamma}^*-I)[u_{n}\Sigma_{t_{-n}\Gamma}^+ u_{n}^*,U_{\Gamma}]\pi_{\Gamma}^+(0) \}  ).
\end{array}\nonumber
\end{equation}
One important observation here is that:
\begin{equation}
u_{\pm n}\Sigma_{t_{\mp n}\Gamma}^\pm u_{\pm n}^*=\mbox{sign}(X_\Gamma^\pm + n),
\end{equation}
(sign$(x)$= the usual sign function) so we can draw the partial conclusion that:
\begin{equation}\label{step1}
\begin{array}{c}
\mbox{Ind}\{\pi_\Gamma^> U_\Gamma \pi_\Gamma^>\}  =-\frac{1}{2} \int d\Gamma 
\mbox{tr}_0\{ (U_{\Gamma}^*-I)  
 [\sum\limits_n \mbox{sign}(X_\Gamma +n),U_{\Gamma}] \}
\end{array}
\end{equation}
As illustrated in Fig.~3, 
\begin{equation}
\sum_n \mbox{sign}(X_\Gamma +n)=S(X_\Gamma)
\end{equation}
 where $S(x)$ is the staircase function shown in Fig.~3. But $S(x)=2x+s(x)$ where $s(x)$ is a bounded periodic function $s(x+n)=s(x)$. Based on this observation, we show that the contribution to the index from $s(X_\Gamma)$ is zero. Indeed, we can follow Refs.~\cite{Prodan:2008oq, Prodan:2008ai} to show that, and under certain circumstances satisfied here, 
 \begin{equation}\label{property}
 \int d \Gamma \mbox{tr}_0 \{ A_\Gamma B_\Gamma \}= \int d\Gamma \mbox{tr}_0 \{B_\Gamma A_\Gamma\}, 
 \end{equation}
 for any covariant operators $A_\Gamma$ and $B_\Gamma$ leaving ${\cal H}_\pm(\Gamma)$ invariant. Since $s(x)$ is bounded, we can open the commutator below,
 \begin{equation}
 \begin{array}{c}
 \int d\Gamma \ \mbox{tr}_0\{ (U_{\Gamma}^*-I)  
 [s(X_\Gamma),U_{\Gamma}] \} \medskip \\
 =\int d\Gamma \ \mbox{tr}_0\{ (U_{\Gamma}^*-I)  
 s(X_\Gamma)(U_{\Gamma}-I) \} \medskip \\
 -\int d\Gamma \ \mbox{tr}_0\{ (U_{\Gamma}^*-I)  (U_{\Gamma}-I)  s(X_\Gamma) \} 
 \end{array}
 \end{equation}
and $s(X_\Gamma)$ is covariant since $s(x)$ is periodic, so due to Eq.~\ref{property} the last two terms cancel each other identically. Thus, we arrived at the conclusion that:
\begin{equation}\label{last}
\mbox{Ind}\{\pi _\Gamma^>U_\Gamma \pi_\Gamma^>\}=-\int d\Gamma \ \mbox{tr}_0\{ (U_{\Gamma}^*-I)[X_\Gamma, U_{\Gamma}] \}.
\end{equation}
But this is exactly Eq.~42 of Ref.~\cite{Prodan:2008ai}, with $\hat{y}_\Gamma$ replaced by $X_\Gamma$. Thus we can repeat the steps of this work to complete our proof (note that property (4) is needed for this). 

\section{Discussion}

Our construction is based on the splitting induced by the time-reversal operation $\theta$: ${\cal H}_\Gamma={\cal H}_-(\Gamma) \oplus {\cal H}_+(\Gamma)$. This splitting is in general not unique, but we introduced several constraints that limit the number of choices. These constraints are: the special requirement (ii) mentioned in Section 3 and the fact that the kernel of the operator $\Sigma^i_\Gamma \equiv \Pi_+^i - \Pi_-^i$ needs to be rapidly decaying. This warrants that our assumption stated before the Main Statement holds true. Since the expectation value of the current is taken only over the states inside the bulk insulating gap, we have to consider only the splitting of the states inside this energy window.

Let us first restrict our discussion to homogeneous edges, in which case the the system is mapped into itself by the discrete translations and the wavenumber $k$ parallel to the  edge is a good quantum number. Dropping the index $\Gamma$, which is no longer needed, the condition (ii) reads: $u_n \Sigma u_n^* = \Sigma$. Thus, the projections $\Pi_\pm^i$ are translational invariant, thus they must be given by sums over the $k$ fibers. For example, the "+" projection must be of the form:
\begin{equation}
\Pi_+^i = \sum_{n_+}\int dk \  |\psi_{n_+,k}\rangle \langle \psi_{n_+,k}|,
\end{equation}
where $\psi_{n,k}$ represent the Bloch functions. The summation goes only over a partial number of band indexes and the integral over $k$ could in principle go only over parts of the Brillouin zone. Note that the phase of the Bloch functions are not relevant here, a good news because the phases are in general difficult to control. Now, since the kernel of this projector, $\Pi_+^i(n_1,n_2;n'_1,n'_2)$, must decay rapidly with the separation $|n_2-n'_2|$, the integral over $k$ must involve the whole interval  $[0, 2\pi]$ and the band $\psi_{n,k}$ must be analytic of $k$. For time reversal invariant spin 1/2 systems, the spectrum is at least doubly degenerate. For the case when the degeneracy is strictly two-fold, or quaternionically simple \cite{Avron:1988kk}, the bands in crystaline systems come in Kramers pairs and one can easily form the projectors $P_\pm^i$ with the required properties. 

\begin{figure}
\center
 \includegraphics[width=7cm]{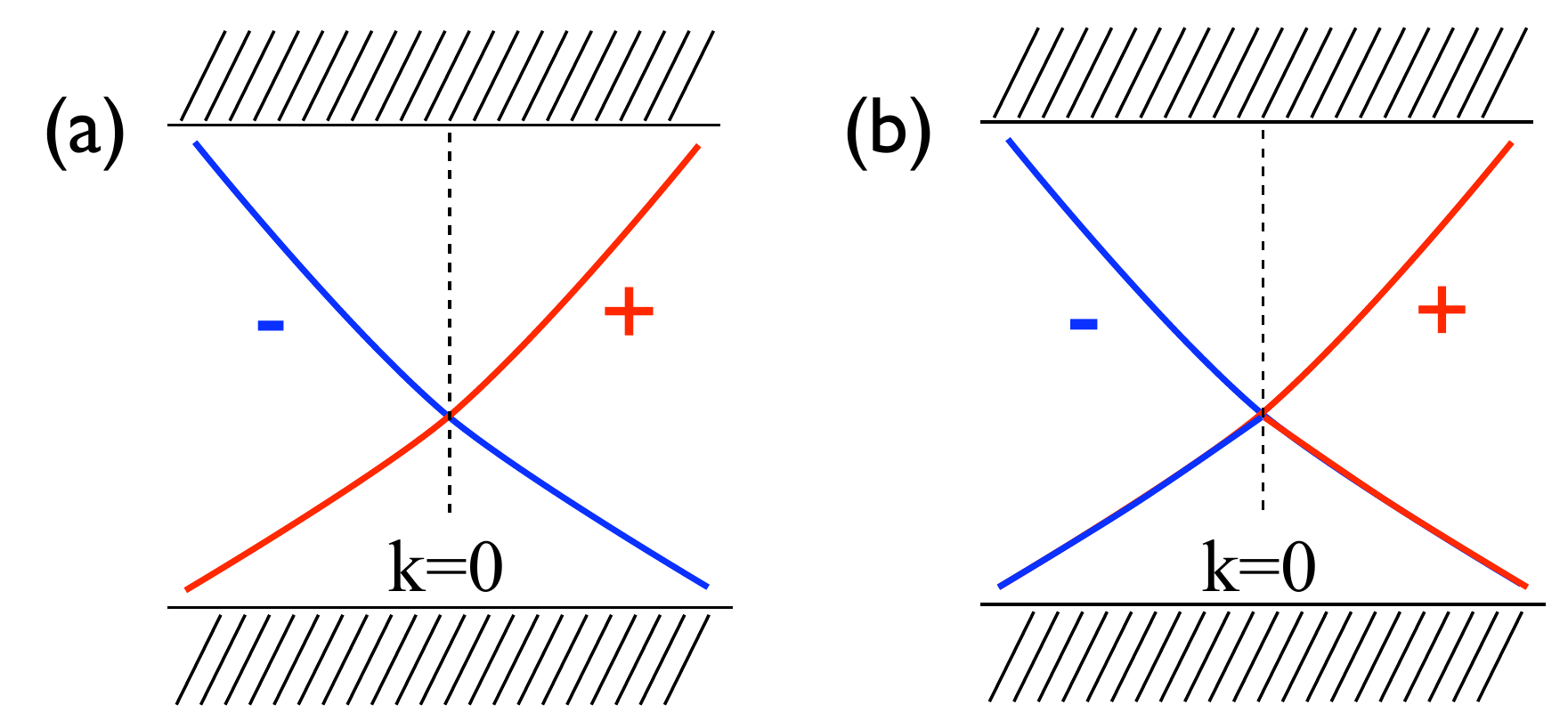}\\
 \caption{The diagram shows the bulk insulating gap and two surface bands crossing it. Both (a) and (b) situations show possible splittings of the states in this energy window into ${\cal H}_\pm$ such that $\theta {\cal H}_+={\cal H}_-$, but only situation (a) leads to projectors $\Pi_\pm^i = \int dk \  |\psi_{\pm,k}\rangle \langle \psi_{\pm,k}|$ decaying exponentially with the separation $|n_2-n'_2|$.}   
 \label{sigma1}
 \end{figure}

Let us exemplify. Consider first the model Eq.~\ref{h1}. As already discussed, there is a pair of Kramers bands crossing the bulk insulating  gap. Fig.~\ref{sigma1}a shows a properly chosen splitting, which gives projectors $P_\pm^i$
that are exponentially decaying with the separation $|n_2-n'_2|$. In contradistinction, Fig.~\ref{sigma1}b shows a bad splitting, which leads to projectors decaying only as $1/|n_2-n'_2|$. As already discussed, the unique choice shown in Fig.~\ref{sigma1}a leads to an edge index equal to 2.

\begin{figure}
\center
 \includegraphics[width=7cm]{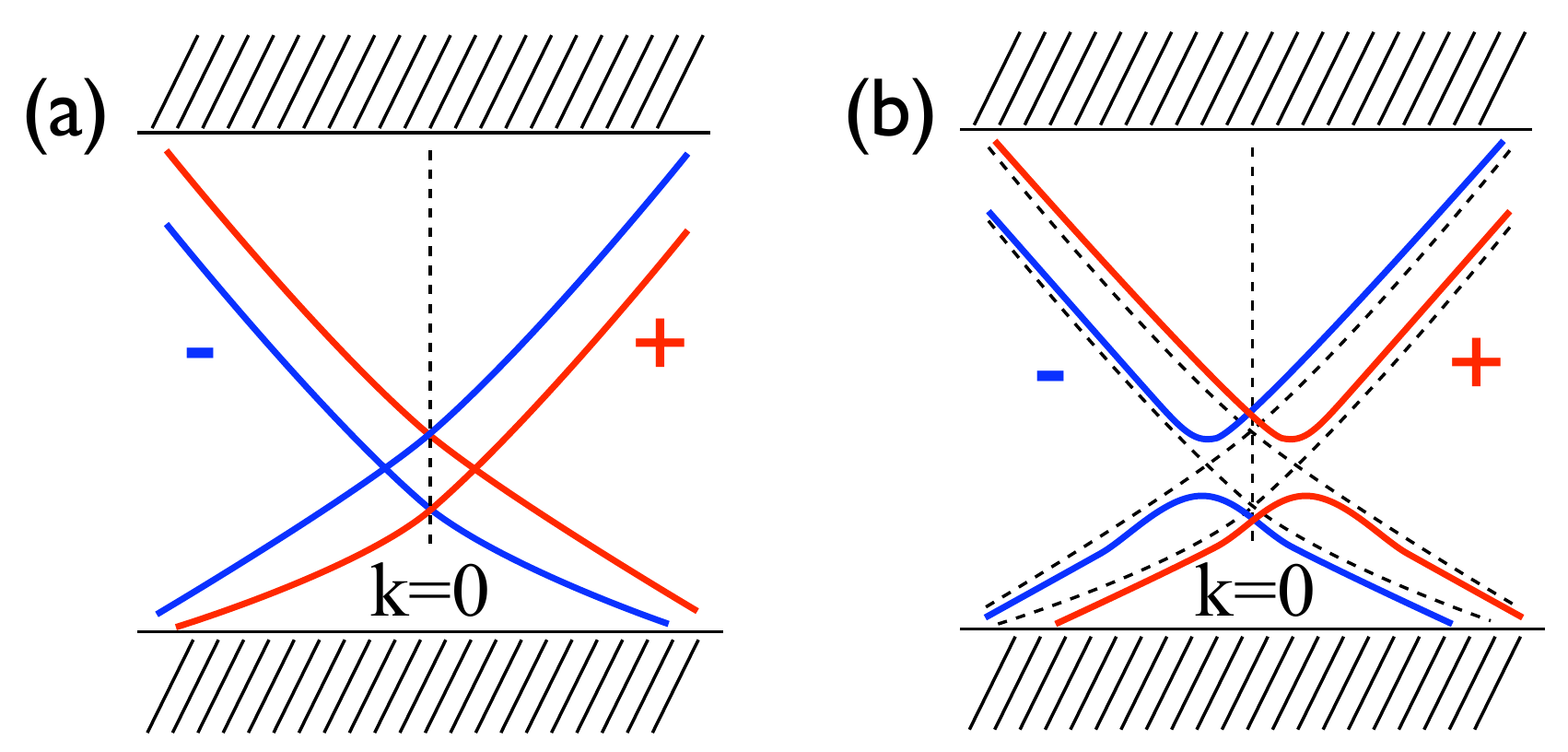}\\
 \caption{The diagram shows the bulk insulating gap and four surface bands crossing it. The case (a) is unstable and small perturbations lead to case (b), which is stable. The diagrams also show the unique splitting of the states in this energy window into ${\cal H}_\pm$ such that $\theta {\cal H}_+={\cal H}_-$, splitting that gives projectors $\Pi_\pm^i = \int dk \  |\psi_{\pm,k}\rangle \langle \psi_{\pm,k}|$ decaying exponentially with the separation $|n_2-n'_2|$ and interpolating smoothly between situations (a) and (b).}  
 \label{sigma2}
 \end{figure}

We consider now a more complex situation in which we have more bands crossing the bulk insulating gap. Let us consider the situation of Fig.~\ref{sigma2}a. This is not quaternionically simple and we know that this case is unstable. The degeneracies at $k$=0 are protected by the time reversal symmetry, but the other two degeneracies will be split by small perturbations. The stable situation is shown in Fig.~\ref{sigma2}b, which is quaternionically simple. In both cases there seems to be more than one possible valid splittings of the states. However, if we want to define the projectors $P_\pm^i$ so that we go continuously (more precise analytically) from situation (a) to situation (b), the splitting can be done in only one way, by incorporating the bands that hybridize when the degeneracies are split into either $P_+^i$ or $P_-^i$. Thus, the only possibility of splitting the bands, for both (a) and (b) situations, is the one depicted in Fig.~\ref{sigma2}. Of course there is a freedom of choice in choosing the $\pm$ labels. With this unique choice, the edge index is zero.

We can continue the argument for more complex situations, but we can already see the general conclusion: the edge index takes only the values 0 or 2, in line with the Z$_2$ topological classification of time-reversal invariant systems. Also, for homogeneous edges, the expectation value of the current of our observable is simply given by the charge current carried by the bands included in the + sector minus the charge current carried by the bands included in the - sector of the $\theta$ splitting. This current is given by an index which is well defined if the $\theta$ splitting leads to a kernel of $\Sigma^i$ that is rapidly decaying with the separation $|n_2-n'_2|$. So the message of our result is that, whenever such $\theta$ splitting exists and the index is non-trivial, there will be edge states that are topologically protected. 

The cases of a rough edge or when a weak bulk random potential is present are more complicated and, at this moment, we can only assume that the projectors $\Pi_\pm^i$ can be properly defined. There is already good progress in characterizing the edge states and computing the Z$_2$ topological invariant for these cases \cite{Essin:wd}. This work adopted an algorithm originally proposed by Fukui and Hatsugai \cite{T:2007yg} for computing the Z$_2$ topological invariant for crystaline systems to the case of non-crystaline systems. We believe that we can adopt this new explicit computational algorithm to construct projectors $\Pi_\pm^i$ with the desired properties, for non-crystaline systems.

\section{Conclusions}

In conclusion, we found that the current of the observable $X=\frac{1}{2}[y \Sigma^i+\Sigma^i y]$ is quantized and that the quantization is given by the index of a Fredholm operator. For the model Eq.~\ref{h1}, this index was shown to take same value as the Spin-Chern number introduced in Ref.~\cite{Sheng:2006vn}. In general, the edge index takes the generic values 0 and 2, in line with the Z$_2$ topological classification of time-reversal invariant systems. 

Our result provides a non-trivial topological invariant that relates directly to the edge of the Quantum Spin-Hall system. The robustness of the edge modes to continuous, time reversal invariant deformations of the model can now be understood from the special properties of the index. We have made a fundamental assumption, namely that the amplitude of $\pi_\Gamma(0)| {\bf n},{\bf a} \rangle$ decays sufficiently fast for large $|n_2|$. For homogeneous edges, we have shown explicitly how to construct $\Sigma^i$ with exponentially decaying kernels, in which case the fundamental assumption holds true. It seems reasonable to assume that one can complete a similar construction for non-homogeneous edges, in which case the analysis shows that the edge conducting channels are robust against random deformations of the edge.\medskip

\noindent {\bf Aknowledgement.} E.P. gratefully acknowledges the hospitality of the Erwin Schrodinger Institute for Mathematical Physics (Vienna) during the summer of 2008. This work was supported by an award from Research Corporation.\medskip

\noindent{\bf References:}\medskip

\bibliographystyle{iopart-num}

\providecommand{\newblock}{}

\end{document}